\documentclass{iaus}
\usepackage{graphicx}

\title
{Rule or exception? \\ Planetary nebulae around hot subdwarf stars}

\author[A. Aller et al.]   
{A. Aller$^1$
, L.F. Miranda$^{1,2}$, A. Ulla$^1$, R. Oreiro$^{3}$, M. Manteiga$^4$,\\E. P\'erez$^1$, C. Rodr\'{\i}guez-L\'opez$^3$}

\affiliation{$^1$Departamento de F\'{\i}sica Aplicada, Universidade de Vigo, Vigo, Spain \\ 
email: {\tt alba.aller@uvigo.es, ulla@uvigo.es, estherperez@edu.xunta.es} \\[\affilskip]
$^2$Consejo Superior de Investigaciones Cient\'{\i}ficas, Madrid, Spain \\email: {\tt lfm@iaa.es} \\[\affilskip]
$^3$Instituto de Astrof\'{\i}sica de Andaluc\'{\i}a - CSIC, Granada, Spain  \\email: {\tt roreiro@iaa.es, crl@iaa.es} \\[\affilskip]
$^4$Departamento de Ciencias de la Navegaci\'on y de la Tierra, Universidade da Coru\~na, \\ A Coru\~na, Spain \\email: {\tt manteiga@udc.es}}

\pubyear{2011}
\volume{283}  
\pagerange{}
\jname{Planetary nebulae: an eye to the future}
\editors{eds. A.\ Manchado, D.\ Sch\"onberner, \& L.\ Stanghellini}
\begin{document}

\maketitle

\begin{abstract}
In this work, we present the first results of an ongoing survey to search for planetary nebulae (PNe) around hot subdwarf stars (sdOs). Deep images and intermediate-resolution long-slit spectra of RWT 152, the only confirmed PN+sdO system in the northern hemisphere, as well as preliminary results for other sdO+PN candidate are presented.
\keywords{(ISM:) planetary nebulae: general, stars: horizontal-branch}
\end{abstract}

\firstsection 
\section{Introduction}

Hot subdwarf O stars (hereafter sdOs) are blue low-mass stars evolving towards the white dwarf phase, although their origin is still unknown. Investigating the association of sdOs with planetary nebulae (PNe) is essential to confirm a post-AGB origin or to favour other progenitors. To date, only four sdOs associated with a PN are known (\cite[Drilling 1983]{Drilling1983}; \cite[Heber \& Drilling 1984]{Heber_Drilling1984}; \cite[Pritchet 1984]{Pritchet1984}), but are they the rule or the exception?  Searches for PN around sdOs were carried out using long-slit spectroscopy and direct images (\cite[M\'endez et al. 1988]{M\'endez_etal88}; \cite[Kwitter et al. 1989]{Kwitter_etal89}) but, unfortunately, these surveys presented some shortcomings regarding the used instrumentation and the small number of observed objects. A more complete survey should be carried out based on deeper images with a larger FoV of a large number of sdOs, and on spectroscopy of the detected extended emission.

We are involved in a survey to search for PNe around sdOs by means of deep images and intermediate-resolution spectroscopy. The results of this survey could contribute significantly to our understanding of the origin and evolution of these puzzling stars.

\section{Observations and Results}

Since July 2010, we have already carried out five observational campaigns for this survey: two with WFC at the Isaac Newton Telescope on Roque de los Muchachos Observatory (La Palma), and three with CAFOS at the 2.2m telescope on Calar Alto Observatory (Almer\'{\i}a). The layout of the observations has been to obtain deep narrow-band [O\,{\sc iii}] and H$\alpha$ (or H$\alpha$+[N\,{\sc ii}]) images of sdOs to search for extended emission around them, complemented with intermediate resolution spectroscopy of the detected cases to confirm the PN 
nature of the nebulae. In order to obtain statistically significant results, we aim to observe both the northern and southern candidate samples. 
 
\begin{figure}[t]
\begin{center}
\includegraphics[width=2.4in]{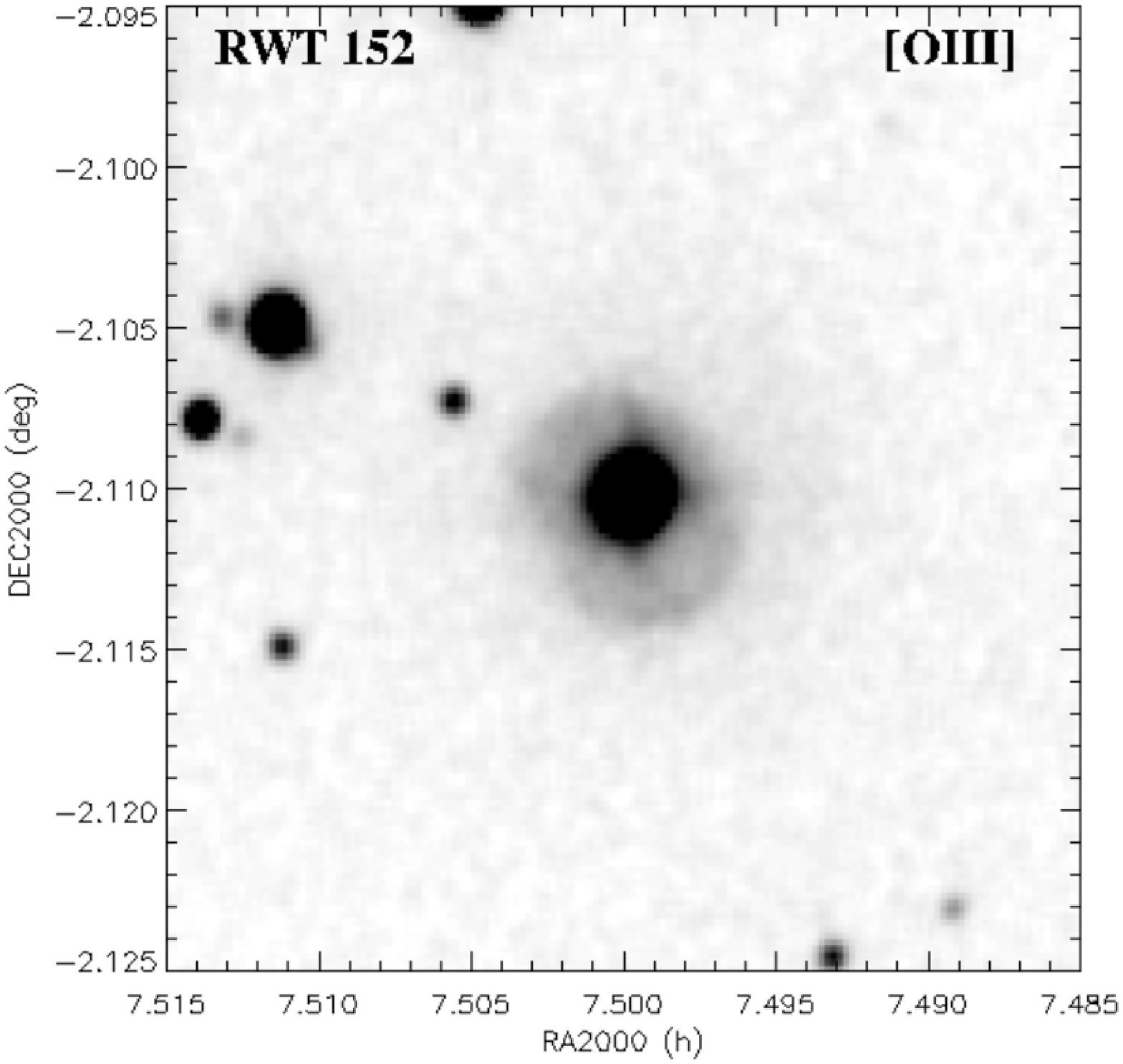} 
 \includegraphics[width=2.85in]{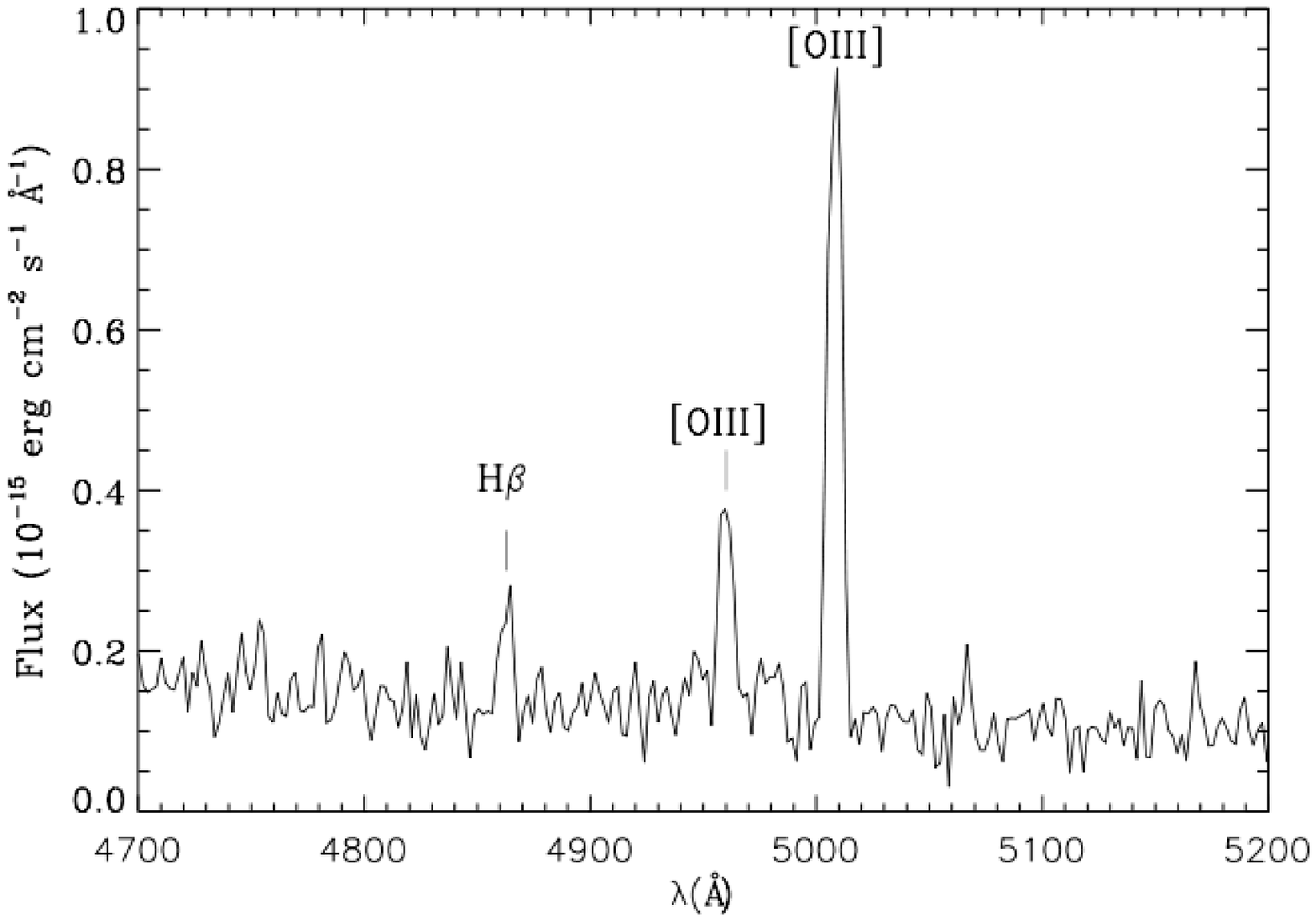} 
 \caption{Left: Grey-scale reproduction of the [O\,{\sc iii}]  image of RWT\,152 obtained with CAFOS. Right: Intermediate-resolution CAFOS spectra of the nebula around RWT\,152 in the 4700-5200 {\AA } spectral range.}
   \label{fig1}
\end{center}
\end{figure}

Among other objects, we observed RWT\,152, the only known PN+sdO system in the northern hemisphere. The [O\,{\sc iii}] image  (Fig.\,\ref{fig1}) shows a non spherical nebula with a size of $\simeq$ 17$''$$\times$21$''$ oriented at position angle $\simeq$ 40$^{\circ}$. In H$\alpha$ (not shown here) the nebula presents a similar appearance to this in [O\,{\sc iii}]. The nebular spectrum  (Fig.\,\ref{fig1}) shows H$\alpha$, H$\beta$ and [O\,{\sc
iii}]$\lambda$$\lambda$4959,5007 line emissions. The [O\,{\sc iii}]/H$\beta$ line intensity ratio is  $\simeq$ 8, a value that is typical of PNe. We analyzed the spectrum of the central using Husfeld et al. (1989) models (kindly provided to us by the authors). Preliminary results for its atmospheric parameters are:  T$_{\rm eff}$ $\simeq$ 46000\,K and log(g) $\simeq$ 4.3, that are similar to those obtained by Ebbets \& Savage (1982) and Chromey (1980).  

 We also detected a new PN candidate around the sdO J19311+4324 (not shown here). H$\alpha$ and traces of H$\beta$ and [O\,{\sc iii}]  emissions are observed in the nebular spectrum but deeper exposures and/or larger telescopes are required to analyze the nebular properties. We plan to continue our program to dilucidate whether sdOs are commonly associated to PNe.
 
 The authors acknowledge financial support from the Spanish MICINN through grants AYA 2009-14648-02, AYA 2008-01934, AYA 2009-08481, and AYA 2010-14840, and from the Xunta de Galicia through grants INCITE09 E1R312096ES, INCITE09 312191PR and IN845B-2010/061, all of them partially funded by FEDER funds.

\end{document}